\documentclass[twocolumn,preprintnumbers,amsmath,showpacs,amssymb,pra,superscriptaddress]{revtex4}

\usepackage{graphicx,epstopdf}
\usepackage{dcolumn}
\usepackage{bm}
\usepackage{amssymb}
\usepackage{amsmath}
\usepackage{color}

\begin{document}

\title{Resonance fluorescence from an asymmetric quantum dot dressed by a bichromatic
electromagnetic field}

\author{$\fbox{G. Yu. Kryuchkyan}\,$}
\affiliation{Institute for Physical Researches, National Academy
of Sciences, Ashtarak-2, 0203, Ashtarak, Armenia}
\affiliation{Yerevan State University, Centre of Quantum
Technologies and New Materials, Yerevan 0025, Armenia}

\author{V. Shahnazaryan}
\affiliation{Science Institute, University of Iceland IS-107,
Reykjavik, Iceland} \affiliation{ITMO University, St. Petersburg
197101, Russia} \affiliation{Institute of Mathematics and High
Technologies, Russian-Armenian (Slavonic) University, Yerevan
0051, Armenia}

\author{O. V. Kibis}\email{Oleg.Kibis(c)nstu.ru}
\affiliation{Department of Applied and Theoretical Physics,
Novosibirsk State Technical University, Novosibirsk 630073,
Russia}\affiliation{Science Institute, University of Iceland
IS-107, Reykjavik, Iceland}\affiliation{Division of Physics and
Applied Physics, Nanyang Technological University 637371,
Singapore}

\author{I. A. Shelykh}
\affiliation{Science Institute, University of Iceland IS-107,
Reykjavik, Iceland} \affiliation{ITMO University, St. Petersburg
197101, Russia}

\begin{abstract}
We present the theory of resonance fluorescence from an asymmetric
quantum dot driven by a two-component electromagnetic field with
two different frequencies, polarizations and amplitudes
(bichromatic field) in the regime of strong light-matter coupling.
It follows from the elaborated theory that the broken inversion
symmetry of the driven quantum system and the bichromatic
structure of the driving field result in unexpected features of
the resonance fluorescence, including the infinite set of Mollow
triplets, the quench of fluorescence peaks induced by the dressing
field, and the oscillating behavior of the fluorescence intensity
as a function of the dressing field amplitude. These quantum
phenomena are of general physical nature and, therefore, can take
place in various double-driven quantum systems with broken
inversion symmetry.

\end{abstract}

\maketitle

\section{Introduction}

Advances in nanotechnology, laser physics and microwave techniques
created a basis for studies of the strong light-matter coupling in
various quantum systems. Differently from the case of weak
electromagnetic field, the interaction between electrons and a
strong field cannot be treated as a perturbation. Therefore, the
system ``electron + strong field'' is conventionally considered as
a composite electron-field object which was called ``electron
dressed by field'' (dressed electron)
\cite{Cohen-Tannoudji_b98,Scully_b01}. The field-induced
modification of physical properties of dressed electrons was
studied in both atomic systems
\cite{Cohen-Tannoudji_b98,Scully_b01,Autler_55} and various
condensed-matter structures, including bulk semiconductors
\cite{Elesin_69,Vu_04,Vu_05}, graphene
\cite{Lopez_08,Kibis_10,Kibis_11_1,Glazov_14,Usaj_14}, quantum
wells
\cite{Mysyrovich_86,Wagner_10,Kibis_12,Teich_13,Shammah,Barachati},
quantum rings \cite{Kibis_11,Kibis_13,Joibari_14,Kyriienko_15},
quantum dots
\cite{Reithmaier,Yoshie,Bloch_2005,Muller_2007,Kibis_09,Ulrich_2011,Majumdar_2012,Savenko_2012,Oster_2012,Paspalakis_2013,Schneider_2015},
etc. Among these structures, quantum dots (QDs) --- semiconductor
3D structures of nanometer scale, which are referred to as
``artificial atoms''
--- seem to be most interesting for optical studies since they are basic elements of modern
nanophotonics \cite{Masumoto_book,Tartakovskii_book}.
 In contrast to natural atoms, the most of QDs are
devoid of inversion symmetry and, therefore, are asymmetric. As an
example, QDs based on gallium nitride heterostructures have a
strong built-in electric field \cite{Widmann_98,Moriwaki_00} and,
therefore, acquire the giant anisotropy
\cite{Williams_05,Bretagnon_06,Warburton_2002,Ostapenko_2010,Ostapenko_2013}.
This motivates to studies of various asymmetry-induced optical
effects in QDs
\cite{Kibis_09,Savenko_2012,Oster_2012,Paspalakis_2013}.

The most of studies of dressed quantum systems was performed
before for a monochromatic dressing field. However, there is a lot
of interesting phenomena specific for quantum systems driven by a
two-mode electromagnetic field with two different frequencies,
polarizations and amplitudes (bichromatic field). In symmetric
quantum systems (atoms and superconducting qubits), the
bichromatic coupling leads to features of photon correlations,
squeezing, Autler-Townes effect, suppression of spontaneous
emission, multi-photon transitions, etc.
 \cite{Ficek_96,Kry1,Kry2,Kry3,Ficek_99,Kry4,Ramirez_2013,Shevchenko_2014}.
Broken symmetry brings substantially new physics to
bichromatically dressed quantum systems, including additional
lines in optical spectra, multiple splitting of the dressed-state
transitions, etc. \cite{Freedhoff_1975,Windsor_1998,Macovei_2015}.
Although these optical effects are extensively studied during long
time, a consistent quantum theory of resonance fluorescence from
bichromatically dressed asymmetric systems was not elaborated
before. The present research is aimed to fill this gap at the
border between quantum optics and physics of nanostructures. To
solve the problem, we focused on the strong light-matter coupling
regime when the interaction of an asymmetric QD with a bichromatic
dressing field overcomes the spontaneous emission and
non-radiative decay of QD excitations. In this case, the spectral
lines of QD are well resolved and various radiation effects
(particularly, the resonance fluorescence) can be analyzed using a
concept of quasienergetic (dressed) electronic states. In the
framework of this approach, such characteristics of dressed QD as
decay rates and lineshapes can be calculated by solving the master
equations in the representation of quasienergetic states. As a
result, we found unexpected features of resonance fluorescence,
which are discussed below.

The paper is organized as follows. In the Section II, we derive
quasienergetic electronic states for an asymmetric QD dressed by a
bichromatic electromagnetic field and calculate matrix elements of
optical dipole transitions between these states. In the Section
III, we apply the found quasienergetic spectrum of dressed
electrons to elaborate the theory of resonance fluorescence from
the QD. The Section IV contains the discussion of the calculated
spectra of resonance fluorescence and the conclusion.

\section{Model of electronic structure}

Let us consider an asymmetric QD with broken inversion symmetry
along the $z$ axis, which is dressed by the bichromatic field
\begin{equation}\label{E}
\mathbf{E}(t)=\mathbf{E}_1\cos\omega_1t+\mathbf{E}_2\cos\omega_2t,
\end{equation}
where the electric field of the first mode, $\mathbf{E}_1$, is
directed along the $z$ axis and the electric field of the second
mode, $\mathbf{E}_2$, is perpendicular to this axis (see Fig.~1a).
In what follows, we will assume that the second frequency,
$\omega_2$, is near the electronic resonance frequency,
$\omega_0$, whereas the first frequency, $\omega_1$, is far from
all resonance frequencies. As a consequence, the bichromatic field
(\ref{E}) mixes effectively only two electron states of QD,
$|1\rangle$ and $|2\rangle$, which are separated by the energy
$\hbar\omega_0$ (see Fig.~1b). Within the basis of these two
states, the asymmetric QD can be described by the matrix
Hamiltonian \cite{Kibis_09}
\begin{figure}[t]
\includegraphics[width=1.0\linewidth]{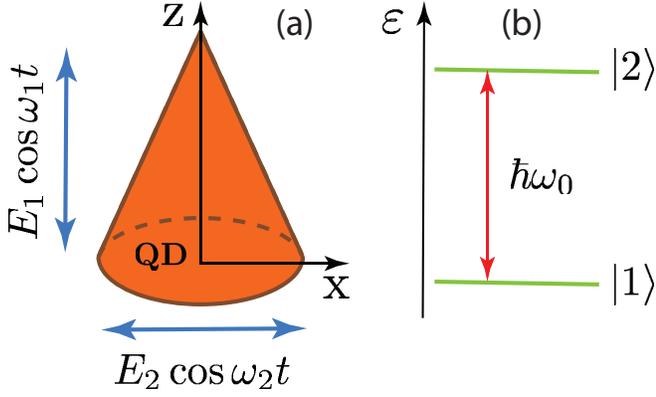}
\caption{(Color online) Sketch of the system under consideration:
(a) Asymmetric quantum dot (QD) with broken inversion symmetry
along the $z$ axis, which is exposed to a bichromatic
electromagnetic field with the electric field amplitudes $E_{1,2}$
and the frequencies $\omega_{1,2}$; (b) Two-level model of the
electronic energy spectrum of the QD, $\varepsilon$, with the
interlevel distance $\hbar\omega_0$.}\label{Fig1}
\end{figure}
\begin{equation}\label{H}
\hat{\cal H}=\begin{pmatrix}
-\hbar\omega_0/2-{d}_{11}{E}_1\cos\omega_1t & -{d}_{12}{E}_2\cos\omega_2t\\
-{d}_{21}{E}_2\cos\omega_2t &
\hbar\omega_0/2-{d}_{22}{E}_1\cos\omega_1t
\end{pmatrix}
\end{equation}
where ${d}_{11}=\langle 1|ez|1\rangle$, ${d}_{22}=\langle
2|ez|2\rangle$ and ${d}_{12}={d}_{21}=\langle 1|ex|2\rangle$ are
the matrix elements of the operator of electric dipole moment
along the $z,x$ axes, and $e$ is the electron charge. To simplify
calculations, the Hamiltonian (\ref{H}) can be written as a sum,
$\hat{\cal H}=\hat{\cal H}_0+\hat{\cal H}^\prime$, where
\begin{equation}\label{dH}
\hat{\cal H}_0=\begin{pmatrix}
-\hbar\omega_0/2-{d}_{11}{E}_1\cos\omega_1t & 0\\
0 & \hbar\omega_0/2-{d}_{22}{E}_1\cos\omega_1t
\end{pmatrix}
\end{equation}
is the diagonal part of the full Hamiltonian (\ref{H}), and
\begin{equation}\label{ndH}
\hat{\cal H}^\prime=\begin{pmatrix}
0 & -{d}_{12}{E}_2\cos\omega_2t\\
-{d}_{21}{E}_2\cos\omega_2t & 0
\end{pmatrix}
\end{equation}
is the nondiagonal part describing electron transitions between
the states $|1\rangle$ and $|2\rangle$ under influence of the
field (\ref{E}). Exact solutions of the non-stationary
Schr\"odinger equation with the Hamiltonian (\ref{dH}),
$$i\hbar\frac{\partial\psi}{\partial t}=\hat{\cal H}_0\psi,$$
can be written in the spinor form as
\begin{equation}\label{psi1}
\psi^{(-)}=e^{i\omega_0t/2}
\exp{\left[i\frac{{d}_{11}{E}_1}{\hbar\omega_1}\sin\omega_1t\right]}
\begin{pmatrix}
1\\0
\end{pmatrix}
\end{equation}
and
\begin{equation}\label{psi2}
\psi^{(+)}=e^{-i\omega_0t/2}\exp{\left[i\frac{{d}_{22}{E}_1}{\hbar\omega_1}\sin\omega_1t\right]}
\begin{pmatrix}
0\\1
\end{pmatrix}.
\end{equation}
Since the two pseudo-spinors (\ref{psi1})--(\ref{psi2}) are the
complete basis of the considered electronic system at any time
$t$, we can sought eigenstates of the full Hamiltonian (\ref{H})
as an expansion
\begin{equation}\label{psi3}
\widetilde{\psi}=a^{(-)}(t)\psi^{(-)} + a^{(+)}(t)\psi^{(+)},
\end{equation}
where the time-dependent coefficients $a^{(\pm)}(t)$ obey the
equation
\begin{eqnarray}\label{a}
i\hbar\dot{a}^{(\mp)}(t)&=&-a^{(\pm)}(t){d}_{12}{E}_2\cos\omega_2te^{\mp
i\omega_0t}\nonumber\\
&\times&\exp\left[\pm
i\frac{({d}_{22}-{d}_{11}){E}_1}{\hbar\omega_1}
\sin\omega_1t\right].
\end{eqnarray}
Applying the Jacobi-Anger expansion,
$$
e^{iz\sin\theta}=\sum_{n=-\infty}^{\infty}J_n(z)e^{in\theta},
$$
we arrive from Eq.~(\ref{a}) at the equation
\begin{eqnarray}\label{a1}
i\hbar\dot{a}^{(\mp)}(t)&=&-a^{(\pm)}(t)\frac{{d}_{12}{E}_2}{2}\sum_{n=-\infty}^{\infty}
J_n\left(\frac{({d}_{22}-{d}_{11}){E}_1}{\hbar\omega_1}\right)\nonumber\\
&\times&\left[e^{\mp i(\omega_0+\omega_2-n\omega_1)t}+e^{\mp
i(\omega_0-\omega_2-n\omega_1)t}\right],
\end{eqnarray}
where $J_n(z)$ is the Bessel function of the first kind. Formally,
the equation of quantum dynamics (\ref{a1}) describes a two-level
quantum system subjected to a multi-mode field. It is well-known
that the main contribution to the solution of such equation arises
from a mode which is nearest to the resonance. Correspondingly,
near the resonance condition,
\begin{equation}\label{res1}
\omega_0\pm\omega_2=n\omega_1,
\end{equation}
we can neglect all modes except the resonant one. In this
approximation, Eq.~(\ref{a1}) reads as
\begin{equation}\label{a1_1}
i\dot{a}^{(\pm)}(t)=a^{(\mp)}(t)F_ne^{\pm i\varphi_n t},
\end{equation}
where $F_n=-({{d}_{12}{E}_2}/{2\hbar})
J_{n}\left(\widetilde{\omega}/\omega_1\right)$ are the Rabi
frequencies of the considered system,
\begin{equation}\label{oo}
\widetilde{\omega}=\frac{E_1(d_{22}-d_{11})}{\hbar}
\end{equation}
is the effective frequency, and
$\varphi_n=\omega_0\pm\omega_2-n\omega_1$ is the resonance
detuning. It follows from Eq.~(\ref{a1_1}) that the considered
problem is reduced to the effective two-level system driven by the
monochromatic field with the combined frequency $\varphi_n$. Using
the well-known solution of Eq.~(\ref{a1_1}) (see, e.g.,
Ref.~[\onlinecite{Landau}]), we can write the sought wave
functions (\ref{psi3}) in the conventional form of quasienergetic
(dressed) states as
\begin{eqnarray}\label{Phi1}
\widetilde{\psi}_1&=&e^{-i \tilde\varepsilon_{1} t/\hbar}\Bigg[\sqrt{\frac12 \left(1+\frac{\varphi_n}{2\Omega_n}\right)} \Lambda_{11}(t)  \begin{pmatrix} 1\\0 \end{pmatrix}\nonumber\\
&-&\sqrt{\frac12 \left(1-\frac{\varphi_n}{2\Omega_n}\right)}
e^{i(\varphi_n-\omega_0)t} \Lambda_{22}(t) \begin{pmatrix} 0\\1
\end{pmatrix}\Bigg],
\end{eqnarray}
\begin{eqnarray}\label{Phi2}
\widetilde{\psi}_2&=&e^{-i\tilde\varepsilon_{2} t/\hbar}\Bigg[\sqrt{\frac12 \left(1+\frac{\varphi_n}{2\Omega_n}\right)} \Lambda_{22}(t)  \begin{pmatrix} 0\\1 \end{pmatrix}\nonumber\\
&+&\sqrt{\frac12 \left(1-\frac{\varphi_n}{2\Omega_n}\right)}
e^{-i(\varphi_n-\omega_0)t} \Lambda_{11}(t) \begin{pmatrix} 1\\0
\end{pmatrix}\Bigg] ,
\end{eqnarray}
where
\begin{equation}\label{psiE}
\tilde\varepsilon_1=-\frac{\hbar\omega_0}{2}-\hbar\Omega_n+\hbar\frac{\varphi_n}{2},\,\,\,\,
\tilde\varepsilon_2=\frac{\hbar\omega_0}{2}+\hbar\Omega_n-\frac{\hbar\varphi_n}{2}.
\end{equation}
are the corresponding quasienergies and
$$\Lambda_{ii}(t)=\exp{\left[i\frac{{d}_{ii}{E}_1}{\hbar\omega_1}\sin\omega_1t\right]},
\,\,\,\,\Omega_n=\sqrt{\frac{\varphi_n^2}{4}+F_n^2}.$$ As to
optical transitions between the dressed states (\ref{Phi1}) and
(\ref{Phi2}), they can be described by the matrix dipole elements
\begin{equation}\label{d1}
\tilde{{d}}_{ij}(t)=\langle
\widetilde{\psi}_i|ex|\widetilde{\psi}_j\rangle=d_{ij}^{(+)} +
d_{ij}^{(-)},
\end{equation}
where $d_{ij}^{(+)}=\left[d_{ji}^{(-)}\right]^*$,
\begin{eqnarray}\label{d}
d_{21}^{(+)}&=&\frac{d_{21}}{2} \left[1+\frac{\varphi_n}{2\Omega_n}\right] \Lambda^*(t) e^{ i(\omega_0-\varphi_n+2\Omega_n)t},\nonumber\\
d_{21}^{(-)}&=&-\frac{d_{12}}{2} \left[1-\frac{\varphi_n}{2\Omega_n}\right] \Lambda(t)   e^{-i(\omega_0-\varphi_n-2\Omega_n)t},\nonumber\\
d_{11}^{(+)}&=&-d_{22}^{(+)}=-\frac{d_{12}}{2}\frac{F_n}{\Omega_n}\Lambda^*(t)e^{i(\omega_0-\varphi_n)t},
\end{eqnarray}
and $\Lambda(t)=\Lambda_{22}(t)\Lambda_{11}^*(t)$.

\section{Resonance fluorescence}

\begin{figure}[t]
\includegraphics[width=1.0\linewidth]{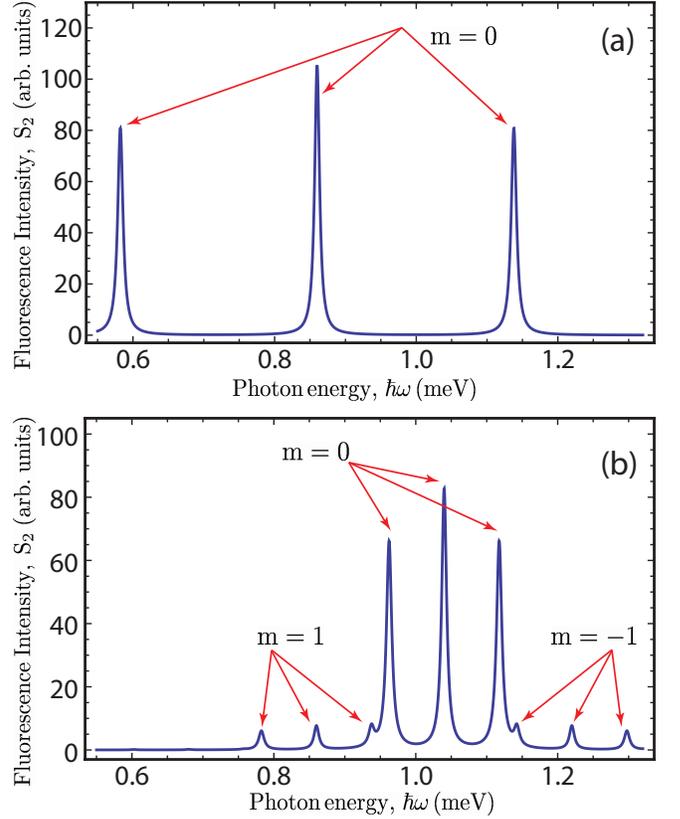}
\caption{(Color online) The spectra of resonance fluorescence from
an asymmetric quantum dot with the interlevel distance
$\hbar\omega_0=1$~meV, dipole moments $d_{22}-d_{11}=d_{12}=40$~D,
and the decay rate $\gamma=10^{-10}$~s$^{-1}$: (a) in the presence
of a monochromatic dressing field with the amplitude $E_2
=3\cdot10^3$~V/cm and the frequency $\omega_2=0.86\,\omega_0$; (b)
in the presence of a bichromatic dressing field with the
amplitudes $E_1 =1.3\cdot10^3$~V/cm, $E_2=3\cdot10^3$~V/cm and the
frequencies $\omega_1=0.18\,\omega_0$, $\omega_2=0.86\,\omega_0$.
The indices $m=0,\pm1$ indicate the different Mollow triplets.}
\label{Fig2}
\end{figure}

The general theory to describe the resonance fluorescence in the
representation of quasienergetic (dressed) states has been
elaborated in Refs.~\cite{Cohen,Scully,Kry5,Kry6}. Applying this
known approach to the considered dressed QD, we have to write the
Hamiltonian of interaction of the QD with the radiative field as
\begin{equation}\label{Hint}
\hat{H}_{\mathrm{int}}=-\tilde{d}_{ij}(t)
[\textbf{E}^{+}(t)+\textbf{E}^{-}(t)],
\end{equation}
where $$\textbf{E}^{\pm}(t)=\int d \omega g^\pm(\omega)
\textbf{e}^\pm_\omega a^\pm_\omega e^{\mp i \omega t}$$ is the
positive (negative) frequency part of the radiative electric
field, $\textbf{E}(t)=\textbf{E}^{+}(t)+\textbf{E}^{-}(t)$, the
parameters $g^\pm(\omega)$, $\textbf{e}^\pm_\omega$,
$a^\pm_\omega$ describe the density, polarization and amplitude of
the corresponding electromagnetic modes, respectively, and
$\tilde{d}_{ij}(t)$ are the dipole matrix elements of dressed QD
(\ref{d1})--(\ref{d}). Within the conventional secular
approximation and Markov approximation \cite{Cohen-Tannoudji_b98},
the equations describing the quantum dynamics of the considered
two-level system read as
\begin{equation}
\begin{aligned}
\frac{d \sigma_{11} (t)}{d t}=-\Gamma_{11}
[\sigma_{11}(t)-\sigma_{11}^S],\,\,\, \frac{d \sigma_{12} (t)}{d
t}=-\Gamma_{12}  \sigma_{12}(t),\nonumber\\
\sigma_{11} (t) + \sigma_{22}(t)=1,
\end{aligned}
\end{equation}
where $\sigma(t)=\mathrm{Tr}\left\{\rho(t)\right\}$ is the reduced
density operator which involves tracing over reservoir variables,
$\sigma_{\alpha\beta}=\langle\widetilde{\psi}_{\alpha}|\sigma|
\widetilde{\psi}_{\beta}\rangle$ are the matrix elements of the
density operator written in the basis of dressed states
(\ref{Phi1})--(\ref{Phi2}),
\begin{equation}\label{SS}
\sigma_{11}^S=\frac{w_{21}}{w_{12}+w_{21}}, \qquad
\sigma_{22}^S=\frac{w_{12}}{w_{12}+w_{21}}
\end{equation}
are the steady-state populations of the dressed states
(\ref{Phi1})--(\ref{Phi2}),
\begin{equation}\label{G}
\Gamma_{11}=w_{12}+w_{21},\,\,
\Gamma_{12}=\frac{\Gamma_{11}}{2}-\mathrm{Re}(M_{11,22}+M_{22,11})
\end{equation}
are the field-dependent decay rates for the dressed states
(\ref{Phi1})--(\ref{Phi2}),
\begin{equation}\label{wij}
w_{12}=\frac{\gamma}{4}\left( 1-\frac{\varphi_n}{2\Omega_n}
\right)^2, \,\,\,\, w_{21}=\frac{\gamma}{4}\left(
1+\frac{\varphi_n}{2\Omega_n} \right)^2
\end{equation}
are the probabilities of radiative transitions per unit time
between the dressed states (\ref{Phi1})--(\ref{Phi2}),
\begin{equation}\label{M}
M_{\alpha,\alpha,\beta,\beta}=\int\limits_0^\infty
\mathrm{d}\tau\, \widetilde{d}_{\alpha\alpha} (t-\tau)
\widetilde{d}_{\beta\beta} (t) \langle E(t) E(t-\tau) \rangle,
\end{equation}
$\langle E(t) E(t-\tau) \rangle$ is the correlation function
averaged over the initial state of electromagnetic field, and
$\gamma$ is the spontaneous emission rate. Substituting
Eqs.~(\ref{wij})--(\ref{M}) into Eqs.~(\ref{SS})--(\ref{G}), we
arrive at the width of the transitions,
\begin{equation}
\Gamma_{11}= \frac{\gamma}{2}\left(
1+\frac{\varphi_n^2}{4\Omega_n^2} \right),\,\,\,\,
\Gamma_{12}=\frac{\gamma}{4}\left(
3-\frac{\varphi_n^2}{4\Omega_n^2} \right),
\end{equation}
and the difference between the populations of dressed electronic
states (\ref{Phi1})--(\ref{Phi2}),
\begin{equation}\label{deltaS}
\Delta_S=\sigma_{11}^S-\sigma_{22}^S=2w_{21}/\Gamma_{11}-1.
\end{equation}
Taking into account the aforesaid, the spectrum of resonance
fluorescence from QD has the form \cite{Scully_b01}
\begin{equation}\label{spint}
S(\omega)\sim\frac{1}{\pi} \mathrm{Re}\left\{
\int\limits_{0}^{\infty} d\tau \langle D^{(+)}(t+\tau)D^{(-)}(t)
\rangle e^{-i \omega \tau} \right\},
\end{equation}
where
\begin{equation}\label{Dmin}
D^{(\pm)}(t)=\sum\limits_{\alpha, \beta} \sigma_{\alpha \beta} (t)
d^{(\pm)}_{\alpha \beta} (t)
\end{equation}
is the positive(negative)-frequency part of the polarization
operator written in the basis of dressed states
(\ref{Phi1})--(\ref{Phi2}). Applying the quantum regression
theorem \cite{Scully_b01} and taking into account
Eqs.~(\ref{d1})--(\ref{d}), we arrive at the correlation function
of the polarization operator (\ref{Dmin}) in the steady-state
regime for long time $t$ and arbitrary time $\tau$,
\begin{equation}\label{1}
\begin{aligned}
&\langle D^{(+)}(t+\tau)D^{(-)}(t)\rangle=\left[
\Delta_S^2+(1-\Delta_S^2)e^{-\Gamma_{11}\tau}\right]\times\\
&\langle d^{(+)}_{11} (t+\tau) d^{(-)}_{11}
(t)\rangle+\frac12\left[(1+\Delta_S)\langle d^{(+)}_{12}
(t+\tau) d^{(-)}_{21}(t)\rangle \right. \\
&+\left.(1-\Delta_S) \langle d^{(+)}_{21} (t+\tau)
d^{(-)}_{12}(t)\rangle \right] e^{-\Gamma_{12}\tau},
\end{aligned}
\end{equation}
where
\begin{equation}
\begin{aligned}
&\langle d^{(+)}_{11} (t+\tau) d^{(-)}_{11} (t)\rangle=\frac{
d_{12}^2 }{4} \frac{F_n^2}{\Omega_n^2}\langle \Lambda^*(t+\tau)
\Lambda(t)\rangle e^{i(\omega_0-\varphi_n)\tau}\\
&\approx\frac{d_{12}^2 }{4}
\frac{F_n^2}{\Omega_n^2}\Sigma(\tau)e^{-i\varphi_n\tau}, \\
&\langle d^{(+)}_{12}(t+\tau)d^{(-)}_{21} (t)\rangle=\frac{
d_{12}^2 }{4} \left(1-\frac{\varphi_n}{\Omega_n}\right)^2
\langle\Lambda^*(t+\tau) \Lambda(t)\rangle\\
&\times e^{i(\omega_0-\varphi_n-2\Omega_n)\tau} \approx \frac{
d_{12}^2 }{4}
\left(1-\frac{\varphi_n}{\Omega_n}\right)^2\Sigma(\tau)
e^{-i(\varphi_n+2\Omega_n)\tau}, \nonumber\\
&\langle d^{(+)}_{21} (t+\tau) d^{(-)}_{12} (t)\rangle = \frac{
d_{12}^2 }{4} \left(1+\frac{\varphi_n}{\Omega_n}\right)^2 \langle
\Lambda^*(t+\tau) \Lambda(t)\rangle\\
&\times e^{i(\omega_0-\varphi_n+2\Omega_n)\tau} \approx \frac{
d_{12}^2 }{4}
\left(1+\frac{\varphi_n}{\Omega_n}\right)^2\Sigma(\tau)e^{-i(\varphi_n-2\Omega_n)\tau},\nonumber\\
\end{aligned}
\end{equation}
and $\Sigma(\tau)=\sum\limits_{m} J_m^2
\left({\widetilde{\omega}}/{\omega_1}\right)
e^{i(\omega_0-m\omega_1) \tau}$. Substituting Eq.~(\ref{1}) into
Eq.~(\ref{spint}), the spectrum of resonance fluorescence,
$S(\omega)=S_1(\omega)+S_2(\omega)$, can be calculated as a sum of
the two parts corresponding to the elastic and inelastic
scattering of light \cite{Scully_b01}, where the elastic term
reads as
\begin{equation}
\begin{aligned}
\label{el_1}
&S_1(\omega)\sim\frac{\Delta_S^2}{\pi}\mathrm{Re}\left\{ \int\limits_0^{\infty} d\tau d_{11}^{(+)}(t+\tau)  d_{11}^{(-)}(t) e^{-i\omega \tau} \right\} \\
&=\left[\frac{ \Delta_S{d_{12}F_n}}{2\Omega_n}\right]^2
\sum\limits_m J_m^2 \left( \frac{\widetilde{\omega}}{\omega_1}
\right) \delta(\omega-[n-m]\omega_1-\omega_2)
\end{aligned}
\end{equation}
and the inelastic term is
\begin{widetext}
\begin{eqnarray}\label{spectrum}
S_2 (\omega)&\sim&\frac{ d_{12}^2 }{4\pi} \left[
\left(1-\Delta_S^2\right)
\left(\frac{d_{12}E_2}{2\hbar\Omega_n}\right)^2J_n^2 \left(
\frac{\widetilde{\omega}}{\omega_1} \right) \sum\limits_m \frac
{J_m^2 \left( {\widetilde{\omega}}/{\omega_1} \right) \Gamma_{11}}
{[\omega-(n-m)\omega_1-\omega_2 ]^2 +\Gamma_{11}^2 } \right.
+\frac{1}{2}\frac{\left(1-{\varphi_n^2}/{4\Omega_n^2}\right)^2}{\left(1+{\varphi_n^2}/{4\Omega_n^2}\right)} \\
&\times&\left.\sum\limits_m \frac {J_m^2
\left({\widetilde{\omega}}/{\omega_1} \right) \Gamma_{12}}
{[\omega-(n-m)\omega_1-\omega_2 +2\Omega_n]^2 +\Gamma_{12}^2
}+\frac{1}{2}
\frac{\left(1-{\varphi_n^2}/{4\Omega_n^2}\right)^2}{\left(1+{\varphi_n^2}/{4\Omega_n^2}\right)}
\sum\limits_m \frac {J_m^2 \left( {\widetilde{\omega}}/{\omega_1}
\right) \Gamma_{12}} {[\omega-(n-m)\omega_1-\omega_2 -2\Omega_n]^2
+\Gamma_{12}^2 } \right].\nonumber
\end{eqnarray}
\end{widetext}
\begin{figure*}[t]
\includegraphics[width=1.0\linewidth]{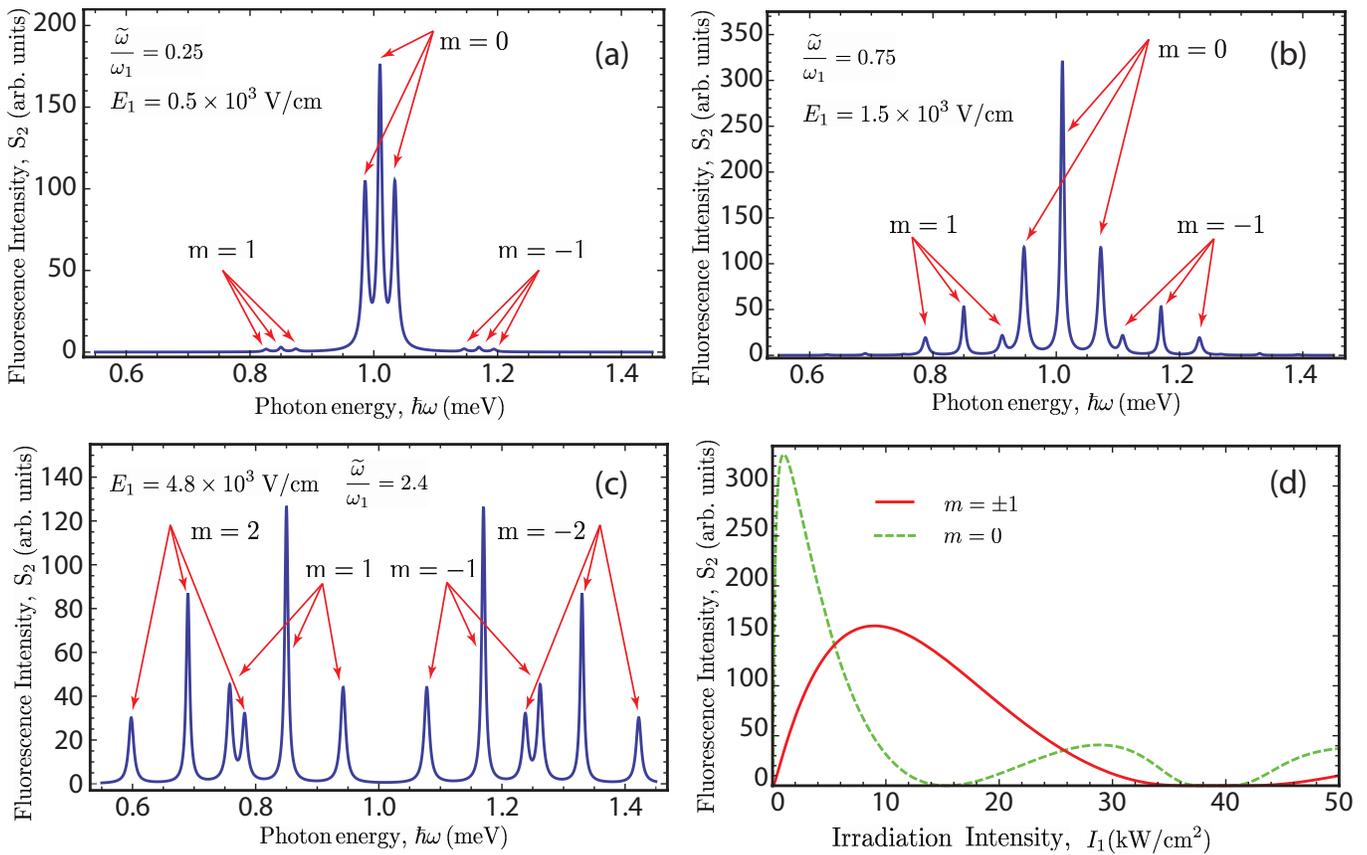}
\caption{(Color online) The spectra of resonance fluorescence from
an asymmetric quantum dot with the interlevel distance
$\hbar\omega_0=1$~meV, dipole moments $d_{22}-d_{11}=d_{12}=40$~D,
and the decay rate $\gamma=10^{-10}$~s$^{-1}$ in the presence of a
bichromatic dressing field with the frequencies
$\omega_1=0.16\,\omega_0$ and $\omega_2=0.85\,\omega_0$, the
amplitude $E_2=2.2\cdot10^3$~V/cm and different amplitudes $E_1$:
(a)--(c) structure of the Mollow triplets with the numbers
$m=0,\pm1,\pm2$; (d) Dependence of the central resonant peaks of
the Mollow triplets with the numbers $m=0,\pm1$ on the irradiation
intensity $I_1=\varepsilon_0E_1^2c/4$.} \label{Fig3}
\end{figure*}
In what follows, we will focus on the inelastic term
(\ref{spectrum}) which is responsible for the spectral features of
the resonance fluorescence.

\section{Discussion and conclusion}

Let us consider the effect of the two key factors of the
considered system --- the bichromatic structure of the dressing
field and the asymmetry of the QD
--- on the resonance fluorescence. Mathematically, these two
factors can be described by the effective frequency (\ref{oo})
appearing in various terms of Eq.~(\ref{spectrum}). In order to
explain it, we have to keep in mind that the ground and excited
states of the asymmetric QD, $|1\rangle$ and $|2\rangle$, do not
possess a certain spatial parity along the asymmetry axis, $z$.
Therefore, the diagonal matrix elements of the dipole moment
operator in an asymmetric QD prove to be nonequivalent,
${d}_{22}\neq{d}_{11}$. As a consequence, the difference of
diagonal dipole matrix elements, ${d}_{22}-{d}_{11}$, describes
the asymmetry of QD \cite{Kibis_09}. Since the effective frequency
(\ref{oo}) is the product of this difference and the first field
amplitude, $E_1$, it can be considered as a quantitative measure
of both the asymmetry of the QD and the bichromatic nature of the
dressing field (\ref{E}). In the following, we will discuss the
dependence of the resonance fluorescence on this effective
frequency, $\widetilde{\omega}=(d_{22}-d_{11})E_1/\hbar$.

The calculated spectra of resonance fluorescence (\ref{spectrum})
are plotted in Figs.~2--3 for different effective frequencies,
$\widetilde{\omega}$, near the resonance (\ref{res1}) with $n=1$.
If the QD is symmetric or the dressing field is monochromatic
($\widetilde{\omega}=0$), the terms with $m\neq0$ in
Eq.~(\ref{spectrum}) vanish since the Bessel functions of the
first kind, $J_m\left({\widetilde{\omega}}/{\omega_1} \right)$,
satisfy the condition $J_m(0)=\delta_{m0}$. The nonzero terms with
$m=0$ correspond physically to the well-known Mollow triplet in
the fluorescence spectrum of a two-level system driven by a
monochromatic field \cite{Scully_b01}, which is plotted in
Fig.~2a. If the QD is asymmetric and the dressing field is
bichromatic ($\widetilde{\omega}\neq0$), the nonzero terms with
$m\neq0$ in Eq.~(\ref{spectrum}) result in the infinite set of
Mollow triplets which can be numerated by the index
$m=0,\pm1,\pm2,...$ (see Fig.~2b and Figs.~3a--3c). It is shown in
Figs.~2--3 that the bichromatic dressing field generates side
Mollow triplets ($m\neq0$), shifts the main Mollow triplet ($m=0$)
and change the amplitudes of the Mollow triplets. According to
Eq.~(\ref{spectrum}), the amplitude of the $m$-th Mollow triplet
is proportional to the squared Bessel function,
$J^2_m\left({\widetilde{\omega}}/{\omega_1} \right)$. This leads
to the oscillating dependence of fluorescence peaks on the
irradiation intensity, $I_1=\varepsilon_0E_1^2c/4$ (see Fig.~3d).
It should be stressed that the zeros of the Bessel function,
$J_m\left({\widetilde{\omega}}/{\omega_1} \right)$, correspond
physically to the zero amplitude of the $m$-th Mollow triplet.
Thus, the dressing field can quench fluorescence peaks.
Particularly, the absence of the main Mollow triplet ($m=0$) in
Fig.~3c is caused by the first zero of the Bessel function
$J_0\left({\widetilde{\omega}}/{\omega_1} \right)$.

Summarizing the aforesaid, we can conclude that the exciting of
asymmetric quantum systems by a bichromatic field results in the
following features of resonance fluorescence spectra: an infinite
set of Mollow triplets, the quench of fluorescence peaks induced
by the dressing field, and the oscillating behavior of the
fluorescence intensity as a function of the dressing field
amplitude. To explain the physics of these novel effects, we have
to stress that the considered bichromatic field (\ref{E}) consists
of an off-resonant dressing field $\mathbf{E}_1$ (which
renormalizes electronic energy spectrum) and a near-resonant field
$\mathbf{E}_2$ (which induces electron transitions between the two
electronic levels). Although the dressing field $\mathbf{E}_1$ is
off-resonant, it is very strong. Therefore, there are noticeable
multiphoton processes which can involve many photons of the field.
Particularly, electron transitions between the electronic levels
can be accompanied by absorption of both near-resonant photon,
$\hbar\omega_2$, and many off-resonant photons, $n\hbar\omega_1$,
where $n$ is the number of the photons. As a consequence, there is
an infinite set of resonances (\ref{res1}) corresponding to the
different numbers $n=0,1,2,...$. Since each resonance is
accompanied with its own Mollow triplet, an infinite set of Mollow
triplet appears in the fluorescence spectrum. As to the
field-induced quench of fluorescence peaks and the oscillating
behavior of the fluorescence intensity as a function of the
dressing field amplitude, these effects arise from the
Bessel-function factors in Eq.~(\ref{spectrum}). Physically, these
factors describe the nonlinear renormalization of electronic
properties with the strong dressing field $\mathbf{E}_1$. It
should be noted that the appearance of the Bessel functions in
expressions describing dressed electrons is characteristic feature
of various quantum systems driven by a dressing field.
Particularly, the similar Bessel-function factors describe
renormalized electronic properties of dressed quantum wells
\cite{Morina_15,Dini_16} and graphene
\cite{Kibis_16,Kristinsson_16}.

Since the considered quantum phenomena depend on electronic
parameters, the elaborated theory paves the new way to the
nondestructive optical testing of various asymmetric structures.
Applying the developed theory to experimental studies of
asymmetric QDs, one should take into account that phonons affect
strongly optical transitions in semiconductor structures. To avoid
the phonon-induced destruction of the discussed fine structure of
the fluorescence spectra, measurements should be performed at low
temperatures, $T$, satisfying the condition
$T\ll\Delta\varepsilon$, where $\Delta\varepsilon=\hbar |F_1|$ is
the characteristic field-induced shift of electron energies (the
dynamic Stark shift).

\begin{acknowledgments}
The work was partially supported by the RISE project CoExAN, FP7
ITN project NOTEDEV, RFBR project 17-02-00053, the Rannis projects
141241-051 and 163082-051, and the Russian Ministry of Education
and Science (project 3.4573.2017). G.Y.K. acknowledges the support
of the Armenian State Committee of Science (project 15T-1C052) and
thanks the University of Iceland for hospitality. O.V.K.
acknowledges the support from the Singaporean Ministry of
Education under AcRF Tier 2 grant MOE2015-T2-1-055. I.A.S.
acknowledges the support of the Russian Government Programm 5-100.

\end{acknowledgments}

\end{document}